\documentclass{article}
\usepackage{graphicx}
\usepackage{amssymb}
\usepackage{cite}
\usepackage[utf8]{inputenc}

\usepackage[T1]{fontenc}

\def\diag{\mathop{\mbox{diag}}}

\begin{document}
\title{RDM-stars and related topics}
\author{Igor Nikitin\\
Fraunhofer Institute for Algorithms and Scientific Computing\\
Schloss Birlinghoven, 53757 Sankt Augustin, Germany\\
\\
igor.nikitin@scai.fraunhofer.de
}
\date{}
\maketitle

\begin{abstract}
In this paper, we will continue to study the model of black holes coupled to the radial flows of dark matter (RDM-stars). According to recent studies, this model well describes the experimental Rotation Curves (RCs) of spiral galaxies, also, RDM-stars can produce signals with the characteristics of Fast Radio Bursts (FRBs). In this paper, we will perform a combined analysis of experimental data on RCs and FRBs, which will allow to constrain tighter parameters of the model. We will also show that within the framework of the model, the Tully-Fisher relation with a slope of $ \beta = 3-4 $ can be obtained. Further, several particular solutions will be considered: tachyonic oven -- a solution in which the incoming flow of tachyons turns into an outgoing flow of massive particles; shell condensate -- a naked singularity of negative mass covered by a thin shell of positive mass, so that the Schwarzschild solution of negative mass inside is joined to one of positive mass outside; also, an RDM solution with a cosmological constant will be considered. Further, in the model under consideration, Penrose diagrams will be constructed, which share common features with Schwarzschild solutions of positive and negative mass, whose combination the model is. The mechanism of effective formation of negative masses in the model is discussed, which is activated when the density of the central core exceeds the Planck value, similarly to the previously studied quantum bounce effect.
\end{abstract}

\section{Introduction}

\begin{figure}
\centering
\includegraphics[width=0.8\textwidth]{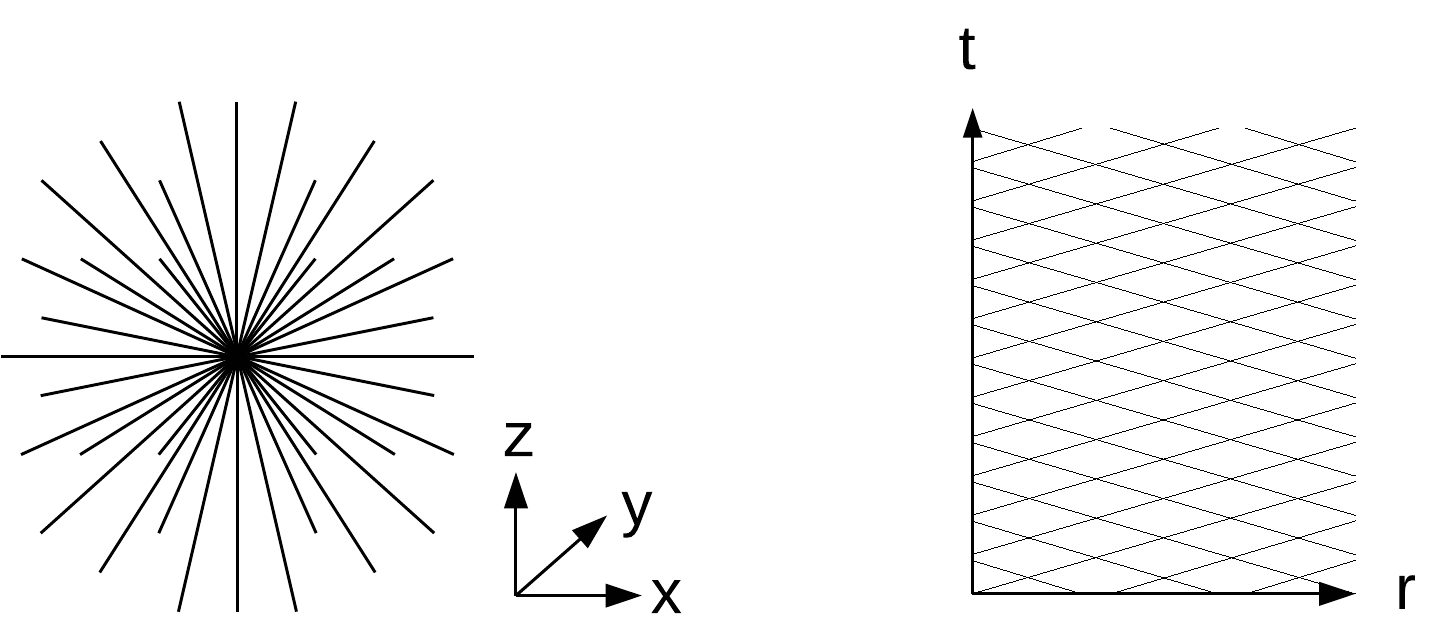}
\caption{RDM-star: a black hole, coupled to radially directed flows of dark matter. Image from~\cite{1701.01569}.}
\label{f0}
\end{figure}

In this paper, we will continue the study of solutions of the general theory of relativity with radial flows of dark matter, the model of RDM-stars, originally formulated in \cite{1701.01569}. This model is shown in Fig.\ref{f0} in $ xyz $ and $ tr $ projections, it consists of incoming and outgoing radially directed flows of dark matter. Flows are energetically balanced, so the solution is stationary.

\begin{figure}
\begin{center}
\includegraphics[width=0.3\textwidth]{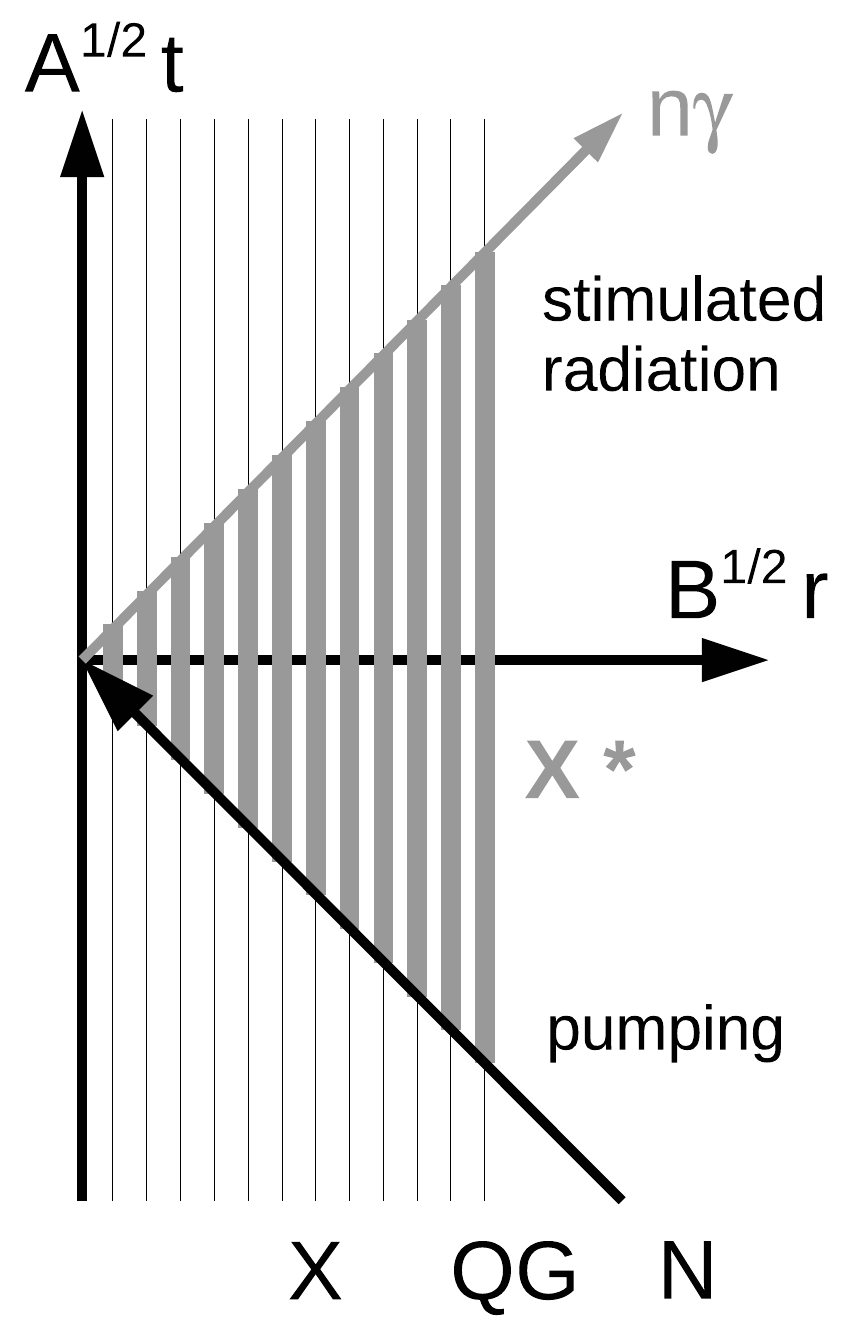}
~~~~\includegraphics[width=0.6\textwidth]{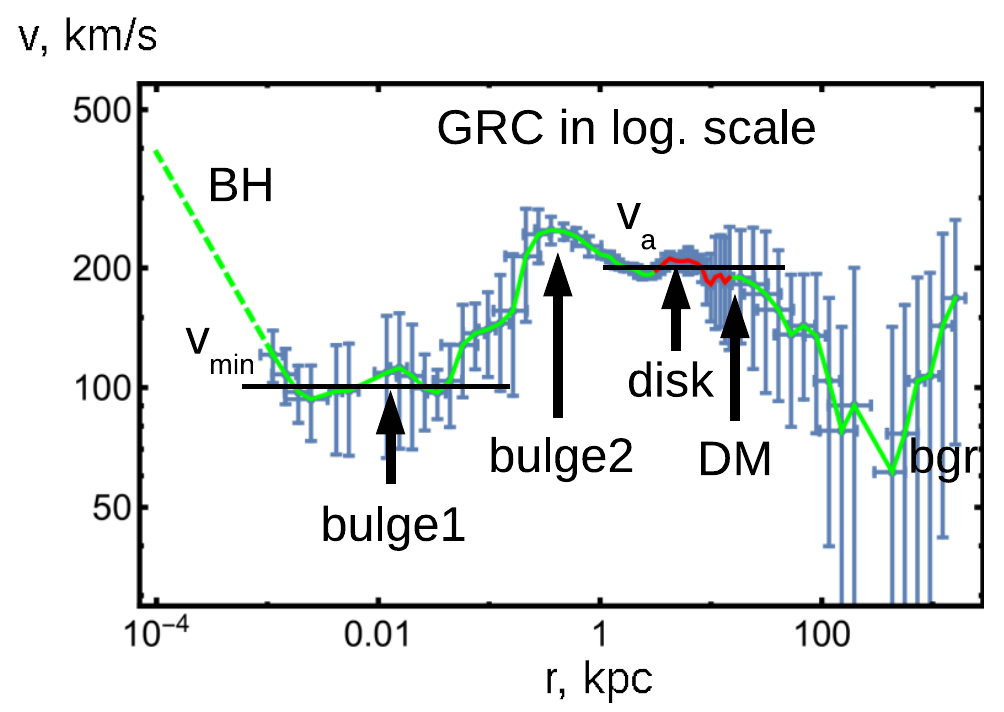}
\end{center}
\caption{Left: FRB generation mechanism in the RDM model. Right: rotation curve for Milky Way galaxy, data from \cite{1307.8241}.}\label{f3}
\end{figure}

In \cite{1707.02764}, solutions of the {\it wormhole} type arising within the RDM model were investigated. Work \cite{1811.03368} investigated the related question of {\it white holes} stability. In work \cite{1812.11801} it was shown that RDM-stars are capable of generating {\it Fast Radio Bursts (FRBs)} with observable characteristics. In work \cite{1903.09972} it was shown that the RDM model well describes the observed {\it Rotation Curves (RCs)} of spiral galaxies.

In this paper, in Section 2, we will perform a combined analysis of FRBs and RCs within the framework of the RDM model and select the range of parameter values that simultaneously describe the observed data for both processes.

In Section 3, we show how, within the framework of the RDM model, the {\it Tully-Fisher relation} can be derived and obtain for this dependence the slope values $ \beta = 3-4 $ with the experimentally observed $ \beta \sim3.6 $.

In Section 4, a particular solution {\it tachyonic oven} will be considered, in which the incoming radial flow is tachyonic and the outgoing is the normal massive matter. This example shows that stationary solutions require only the energy balance of incoming and outgoing flows and are not necessarily T-symmetric.

Section 5 will consider the solution {\it shell condensate}. To construct this solution, we consider the RDM model of a massive type with two turning points. It will be shown that in the limit when these two points approach each other, an interesting stationary solution arises, in the form of a thin shell of positive mass in equilibrium in the field of the central singularity of negative mass. The total mass of this configuration is positive, and from the outside the system is identical to the black hole of the Schwarzschild type.

In Section 6, the modification of the solution will be investigated when {\it the cosmological constant} is introduced into equations.

In Section 7, we exercise in constructing {\it Penrose diagrams} in the RDM model and compare them with Penrose diagrams for Schwarzschild solutions of positive and negative mass, whose combination the model is.

Section 8 will discuss the mechanism for the formation of negative masses, which is activated when the Planck density is exceeded and is similar to the {\it quantum bounce effect}.

\section{Combined analysis of FRBs and RCs in RDM model}

According to \cite{1812.11801}, the RDM model can describe FRBs in the following scenario, see Fig.\ref{f3} on the left. The object of an asteroid mass falls on an RDM-star. The superstrong gravitational field present inside the RDM-star acts as a super-power accelerator, which boosts the nucleons composing the asteroid to ultrarelativistic energies. In the depths of the RDM-star the core is located, consisting of particles of Planck mass. Accelerated nucleons enter into inelastic scattering reactions with Planck particles and transfer them to the excited state (pumping). The decay of excited states generates photons that cause subsequent transitions through the standard mechanism of stimulated radiation (laser). The outgoing short pulse of coherent radiation turns out to be strongly collimated due to the special configuration of the gravitational field, which aligns the light geodesics in the radial direction. On the way out, the impulse is experiencing a superstrong gravitational redshift, which transfers it from ultrahigh frequencies to radio band. The calculation in \cite{1812.11801} shows that the frequency distribution of the obtained FRB is bounded from above. The low-frequency signals have a large scatter broadening, reducing the amplitude, with the result that the registration of signals at the cutoff frequency may be preferable. The cutoff frequency is determined by the formula
\begin{eqnarray}
&\nu_{out}=2^{-5/4}\pi^{-1/4}c\, r_s^{-1/2}\lambda_N^{-1/2}\epsilon^{1/4},\label{nuout1}
\end{eqnarray}
where $ \lambda_N $ is the Compton wavelength of the nucleon, $ r_s $ is the gravitational radius of the RDM-star, $ \epsilon $ is the parameter determining the density of dark matter, $ c $ is the speed of light. The parameter $ \epsilon $ is related to the observed asymptotic orbital velocity $ v_a $ of objects rotating around the RDM-star, by the ratio $ \epsilon = (v_a / c) ^ 2 $. Considering the Milky Way galaxy (MW) and attributing the entire distribution of dark matter in it to the central black hole, we obtain for the parameters
\begin{eqnarray}
&\lambda_N=1.32\cdot10^{-15}\textrm{m},\ r_s=1.2\cdot10^{10}\textrm{m},\ v_a=200\textrm{km/s},\ \epsilon=4\cdot10^{-7},
\end{eqnarray}
the frequency value
\begin{eqnarray}
&\nu_{out}=600\textrm{MHz},
\end{eqnarray}
located within the observed FRB range 111MHz-8GHz \cite{ATel11932,ATel11901,1804.04101}.

In work \cite{1903.09972}, the model of RDM-stars was used to describe the fine structure of galactic rotation curves. It was assumed that not only the central, but all black holes in the galaxy are the sources of dark matter flows. The distribution of stellar black holes in the galaxy was assumed to be proportional to the density of luminous matter, for which, in turn, the generally accepted models were chosen (exponential disk \cite{Freeman1970}, exponential spheroid \cite{1307.8241}). A comparison was made of predictions of the model with two types of experimental rotation curves: the Universal Rotation Curve (URC, \cite{9506004}) is the result of averaging 1000+ galaxies, leading to the smoothing of individual structures (minima, maxima); Grand Rotation Curve (GRC, \cite{1307.8241}), a rotation curve for the Milky Way, measured in a large range of distances. For both types of experimental curves, a good fit with the RDM model was obtained. Since for URC the experimental $ v (r) $ curves are normalized to values at the optical radius $ (r / R_{opt}, v / v_{opt}) $, and in this work we are interested in absolute values, we concentrate on GRC, where such normalization is not performed. When analyzing the GRC in \cite{1903.09972}, three scenarios are used, that are distinguished by the coupling coefficients of dark matter with individual galactic structures: the central black hole, the inner bulge, the outer bulge and the disk. The contribution to the parameter $ \epsilon $ defined above for the central black hole is given by the formula
\begin{eqnarray}
&\epsilon_{smbh}=GM_{smbh}\lambda_{smbh}/(c^2L_{KT}),
\end{eqnarray}
where $ G $ is the gravitational constant, $ M_{smbh} $ is the mass of the central supermassive black hole, $ L_{KT} $ is the constant of the unit of length defined in the fit, $ \lambda_{smbh} $ is the fixed coupling constant corresponding to the central black hole. The parameters $ L_{KT} $ and $ \lambda_{smbh} $ for the three considered scenarios and the corresponding values of the parameter $ \epsilon_{smbh} $ are given in Table~\ref{tab1}. The absolute maximum of the parameter $ \epsilon_{smbh} = 1.1 \cdot10 ^{- 7} $ corresponds to the absolute minimum speed on GRC, which is $ v_{min} = 100 $km/s. Fig.\ref{f3} on the right shows the GRC from the data from \cite{1307.8241}, along with the characteristic values of $ v_{min} $ and $ v_a $.

\begin{table}
\begin{center}
\caption{GRC, 3 scenarios, coupling coefficients and fitting results}\label{tab1}

~

\def\arraystretch{1.1}
\begin{tabular}{|c|c|c|c|}
\hline
$par$&s1&s2&s3\\
\hline
 $\lambda_{smbh}$& 0& 1& $10^3$ \\
 $\lambda_1$& 0& 1& $10^2$ \\
 $\lambda_2$& 0& 1& 2 \\
 $\lambda_{disk}$& 1& 1& 1 \\ \hline
 $R_{opt}$/kpc& $7.7$ & $8.0$ & $9.0$ \\
 $L_{KT}$/kpc& $7.0$ & $6.3$ & $12.0$ \\
 $R_{cut}$/kpc& $58$ & $45$ & $53$ \\ \hline
 $\epsilon_{smbh}$& $0$ & $3.1\cdot10^{-11}$ & $1.6\cdot10^{-8}$ \\
 $x=L_{KT}/R_{cut}$& $0.12$ & $0.14$ & $0.23$ \\
\hline

\end{tabular}

\end{center}
\end{table}

For stellar black holes of mass $ M_{sbh} \sim10M_ \odot $, the same contribution to the $ \epsilon $ parameter was assumed, that is, the total contribution was determined by the number of black holes, without taking into account their individual characteristics. Good quality of the fit confirms this assumption. For the total number of stellar black holes in our galaxy there is a broad estimate of $ N_{sbh} = 10 ^ 6-10 ^ 9 $, see \cite{1107.3165} and references therein. Thus, we can estimate the contribution of each black hole from this class as $ \epsilon_{sbh} = (\epsilon- \epsilon_{smbh}) / N_{sbh} $, for $ \epsilon = 4 \cdot10 ^{- 7 } $.

\begin{figure}
\begin{center}
\includegraphics[width=0.6\textwidth]{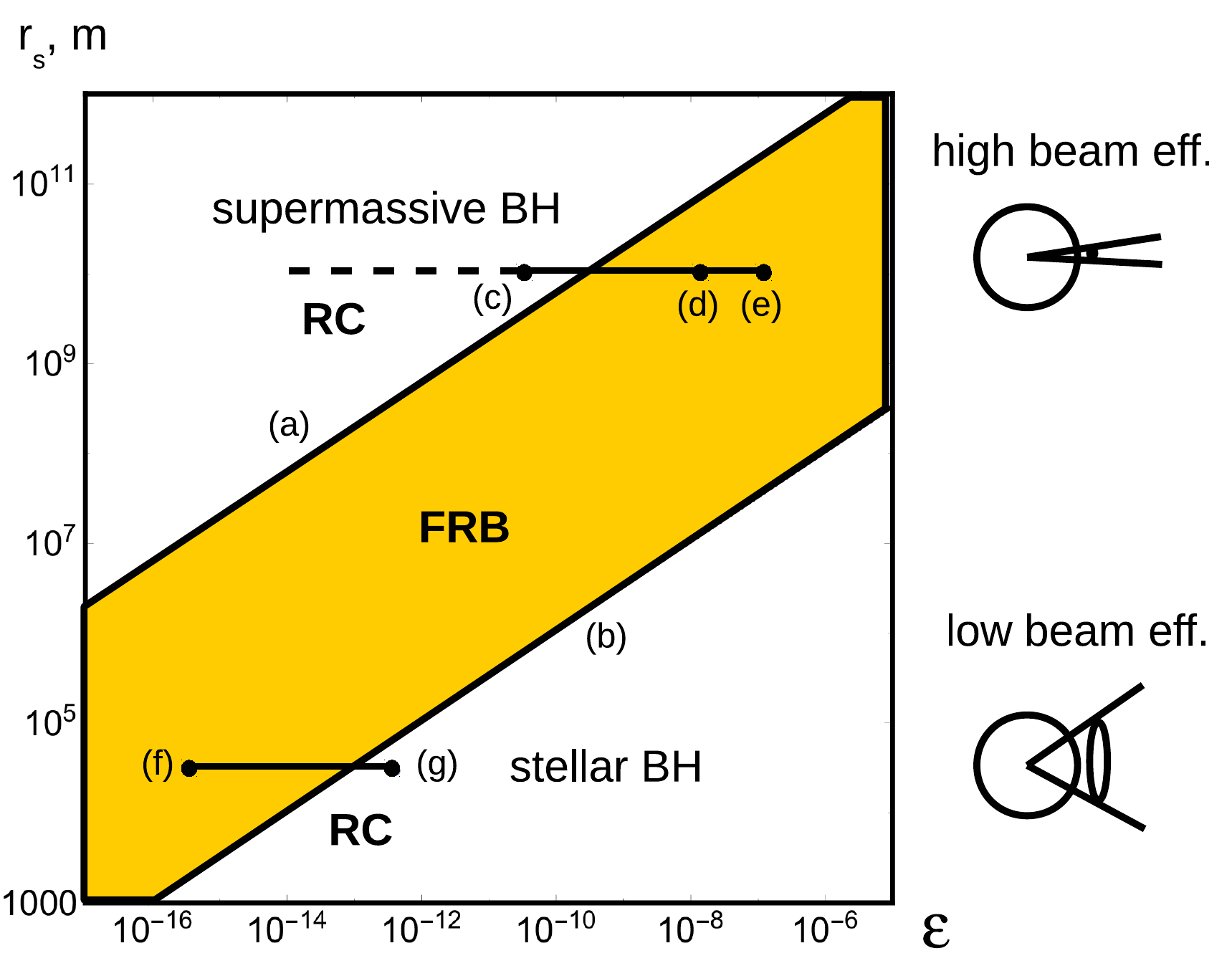}
\end{center}
\caption{Combined analysis of RCs and FRBs.}\label{f4}
\end{figure}

\begin{figure}
\begin{center}
\includegraphics[width=0.45\textwidth]{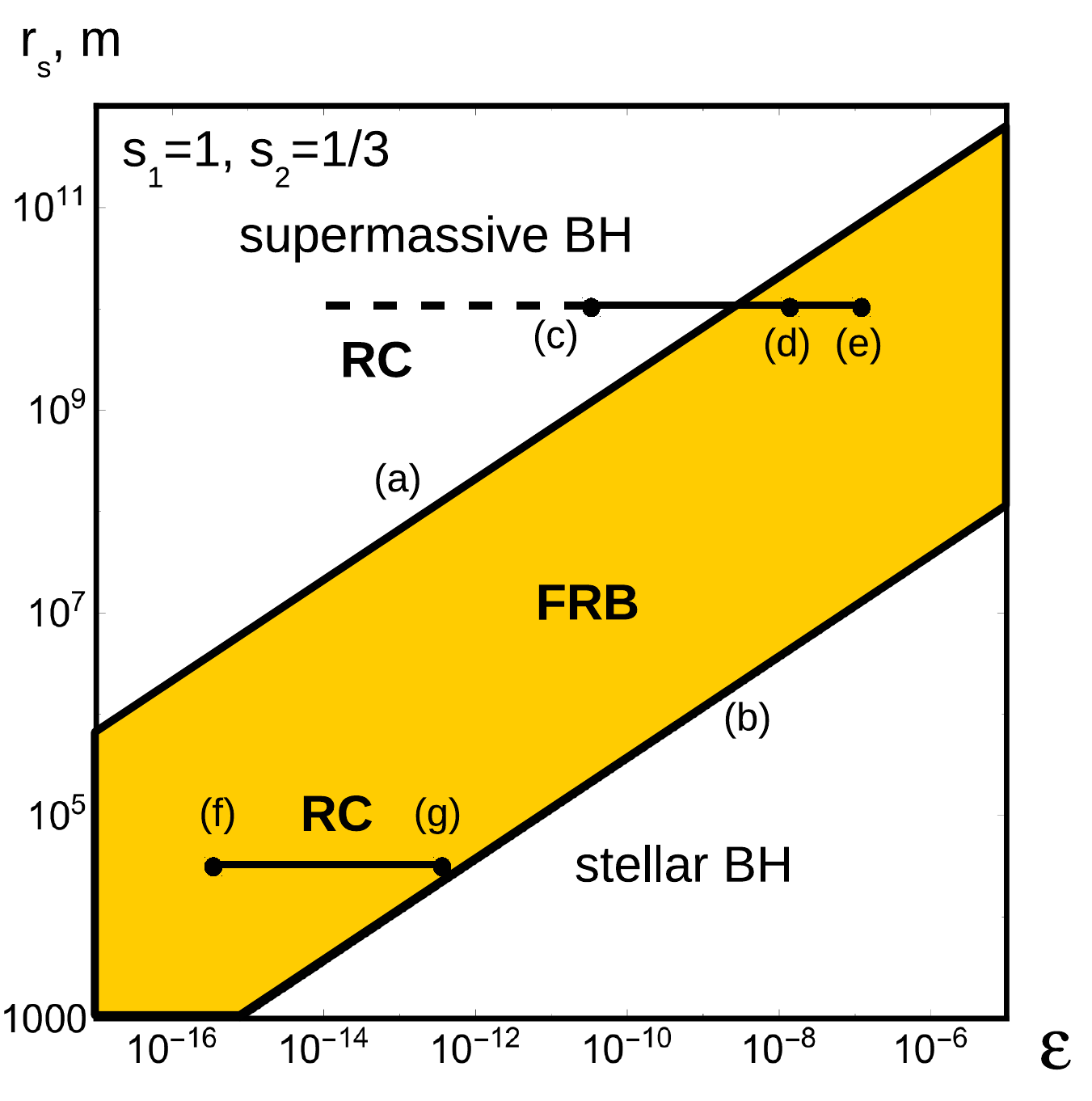}
~~~~\includegraphics[width=0.45\textwidth]{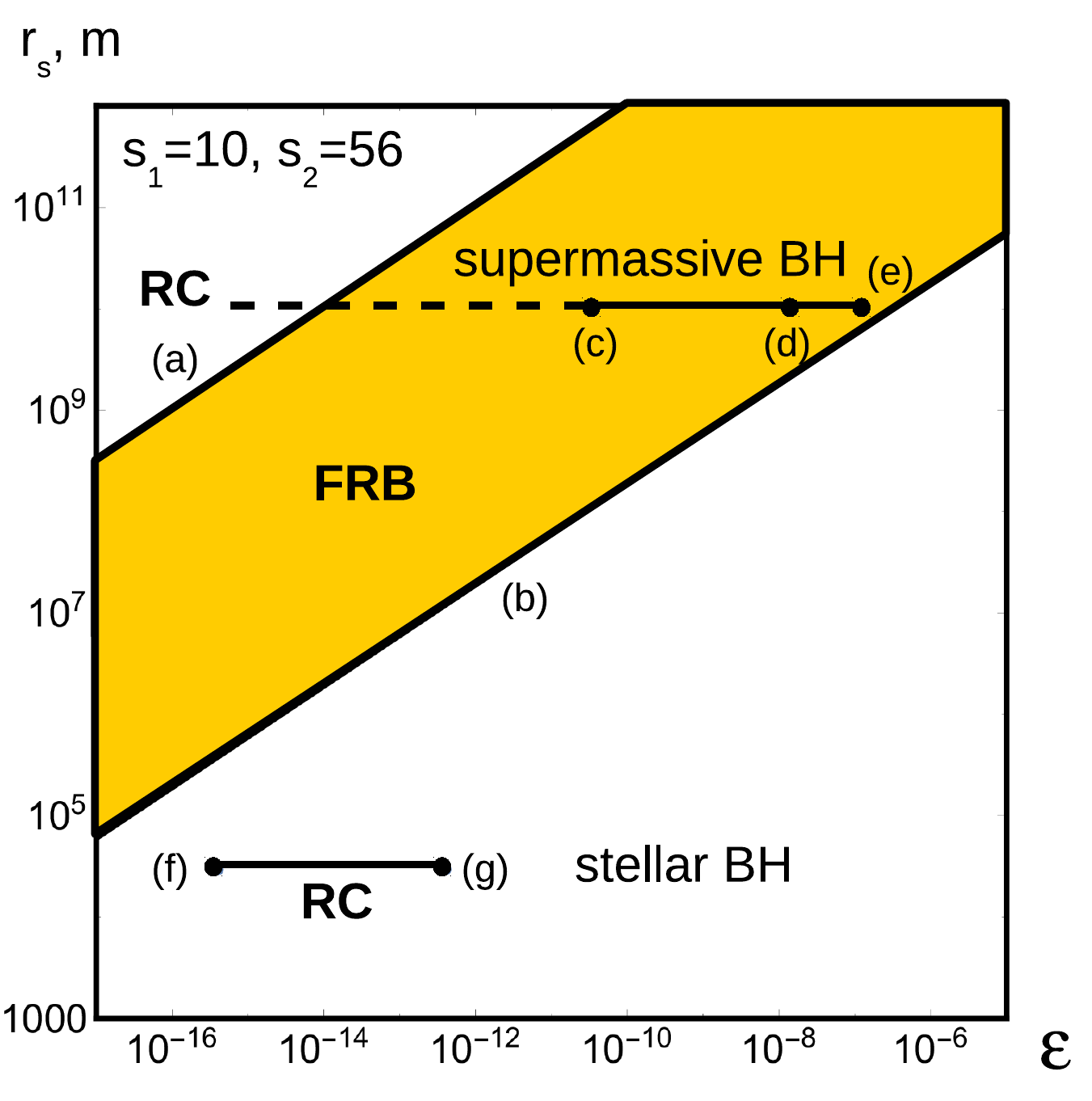}
\end{center}
\caption{The same as above, with the variation of FRB tuning parameters $ s_{1,2} $.}
\label{f5}
\end{figure}

In Fig.\ref{f4}, the results of the analysis of FRBs and RCs are combined. The band corresponds to the experimentally observed range of FRBs according to formula (\ref{nuout1}), (a) $ \nu_{out} = 111 $ MHz, (b) $ \nu_{out} = 8 $ GHz. The horizontal lines show the estimates obtained in the analysis of GRC: for the central supermassive black hole: (c) scenario s2, (d) scenario s3, (e) absolute maximum $ \epsilon_{smbh} $ for MW; for the stellar black holes: (f) the estimate for $ N_{sbh} = 10 ^ 9 $, (g) the same for $ N_{sbh} = 10 ^ 6 $, assuming $ \epsilon_{smbh} \ll \epsilon $.

First of all, we see that the intersection of the experimental characteristics under consideration exists, which means that the model is capable of describing at least a subset of FRBs as a result of an asteroid falling on the core of the RDM-star. There are two classes of solutions, the sources of FRBs can be both central supermassive and stellar black holes. As noted in \cite{1812.11801}, supermassive sources are more preferable. Since they have much larger radius, they have a smaller beam width and greater energy transfer efficiency over long distances, which makes their registration more preferable. Also, dense plasma surrounding supermassive black holes can create a large scatter broadening, characteristic for the observed signals. In works \cite{1401.1795,1512.00529}, the nuclei of other galaxies were also considered as the preferred sources of FRBs.

Work \cite{1812.11801} also introduced FRB tuning parameters that modify the formula for the cutoff frequency as
\begin{eqnarray}
&\nu_{out}=s_1^{1/4}s_2^{1/2}2^{-5/4}\pi^{-1/4}c\, r_s^{-1/2}\lambda_N^{-1/2}\epsilon^{1/4},
\end{eqnarray}
where $ s_1 = 1 ... 10 $ is a tuning parameter that simulates earlier onset of QG effects, activating them when the density is $ s_1 $ times less than the Planck one; $ s_2 = 1/3 ... 56 $ is a fragmentation factor that regulates the mass of falling particles from the constituent quarks to iron nuclei. Taking into account these parameters, the band on the graphs is shifted, Fig.\ref{f5} left for the minimum, right for the maximum values of FRB tuning parameters. After this variation, the intersection of the FRBs and RCs covers almost the entire band of observed frequencies. In particular, for the preferred scenario of supermassive black holes, only a narrow zone $ \nu_{out} = 6.33 ... 8 $GHz remains uncovered, and this gap will close if $ r_s $ is a factor of 1.6 smaller or $ \epsilon_{ smbh} $ is a factor of 2.6 larger. Such variations of parameters can be attributed to the natural variability of the source, as well as the considered variations in the parameters $ s_{1,2} $ and the coupling constants in the scenarios (c), (d), ...

To some extent, our analysis was related to the MW parameters extracted from its rotation curve. Thus, we characterize a subset of FRBs, whose sources are located in galaxies similar to MW by their parameters. For the version with supermassive FRB source, the key parameters are the asymptotic orbital velocity $ v_a $, the gravitational radius of the source $ r_s $ and the fraction of the parameter $ \epsilon = (v_a / c) ^ 2 $, which can be attributed to the central source. For the stellar black holes, the characteristic gravitational radius is common \cite{1006.2834}, the key parameter is the fraction of the parameter $ (\epsilon- \epsilon_{smbh}) / N_{sbh} $, which can be attributed to every stellar black hole in the galaxy.

It would be interesting to populate this diagram with data from other galaxies. For this, however, it is necessary that the galactic RC be resolved till the central black hole. This is required for the detailed fit or at least $ v_{min} $ estimate to find the contribution of the central black hole to the $ \epsilon $ parameter. With insufficient resolution, the contributions of smbh and bulge1,2 are accumulated in one term. As a result, all black holes contained in the central region (supermassive, intermediate mass, stellar) effectively gather into one large black hole. FRB, however, occurs on distinct black holes, which requires an assessment of their individual parameters.

\section{Derivation of Tully-Fisher relation in RDM model}

In this section, we show how, within the framework of the RDM model, the Tully-Fisher relationship can be derived:
\begin{equation}
M_{lm}\sim v_a^\beta,
\end{equation}
where $ M_{lm} $ is the mass of luminous matter (which in this work is identified with the total baryonic mass, it includes the masses of stars, interstellar gas, dust nebulae, etc.), $ v_a $ is the asymptotic orbital velocity of the circular motion of objects at a large distance from the center of the galaxy, the slope index is $ \beta = 3.64 \pm0.28 $ \cite{1106.0505}.

There is a similar calculation in \cite{0604496}, which we adapt to our model. For a compact distribution of luminous matter, the mass of dark matter associated with it in a sphere of radius $ R $ and the asymptotic orbital velocity at a large distance are expressed by the formulas
\begin{equation}
M_{dm}(R)=M_{lm} R/L_{KT},\ v_a^2=GM_{dm}(R)/R=GM_{lm}/L_{KT}.
\end{equation}
Further, a distance $ R_{cut} $ is introduced at which the RDM model is cut and joined with a uniform distribution of matter. The junction is done as follows, \cite{1903.09972}. According to the GRC behavior in Fig.\ref{f3} on the right, immediately after the $ R_{cut} $ the Kepler term dominates, corresponding to the total mass of the galaxy in the $ R_{cut} $ sphere, with the result that the curve has a dip. Then, the contribution of the uniformly distributed matter bgr starts to dominate, which has a {\it local overdensity} of the luminous and dark matter with respect to the cosmological values. Due to this overdensity, the bgr contribution to GRC turns out to be growing (attractive), in contrast to the repulsive behavior of uniformly distributed matter with the dominant cosmological term. The curve is traced to distances of 1.6Mpc, of the order of the radius of the Local Galactic Group, up to which the bgr curve is growing.

\begin{figure}
\begin{center}
\parbox{0.4\textwidth}{\includegraphics[width=0.3\textwidth]{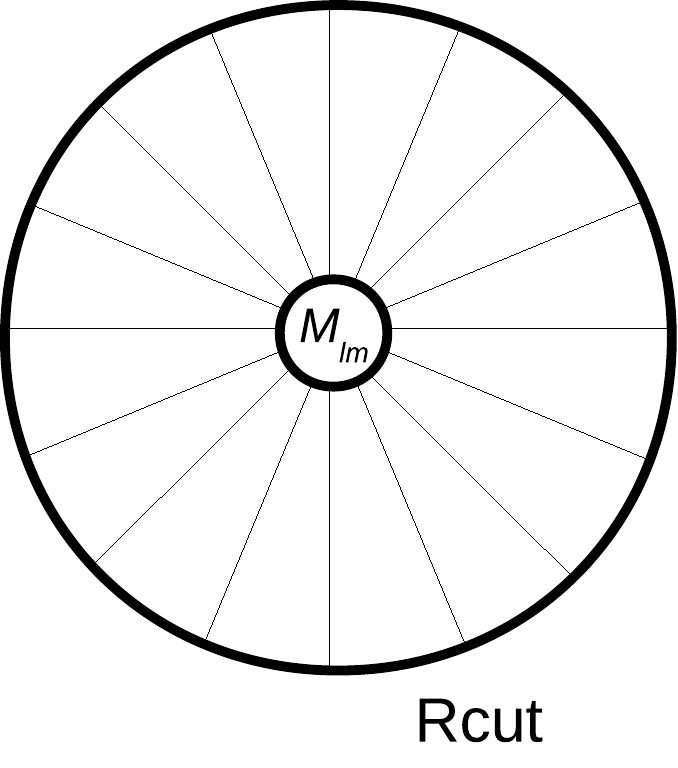}}
~~~~\parbox{0.5\textwidth}{\caption{Illustration to the derivation of Tully-Fisher relation. When a galaxy is formed, the luminous matter from $ R_{cut} $ sphere collapses to the center, while the dark matter from the same sphere goes to the RDM configuration.}}
\end{center}
\label{f6}
\end{figure}

Further, we assume that during the formation of galaxies, a collapse of the initially homogeneous luminous matter occurs in a sphere of radius $ R_{cut} $ to distances of the order of $ R_{opt} \ll R_{cut} $, while the dark matter within $ R_{cut} $ sphere transforms to the RDM configuration. The result of this process is shown in Fig.\ref{f6}. In this case, the ratio of the contributions of the luminous and dark matter in $ R_{cut} $ sphere is comparable with their cosmological ratio. This was shown by the calculation in \cite{1903.09972}, which can be easily reproduced here:
\begin{equation}
x=M_{lm}/M_{dm}(R_{cut})=L_{KT}/R_{cut}\approx\Omega_{lm}/\Omega_{dm}=0.19, 
\end{equation}
where we used cosmological values from \cite{1807.06209}. In \cite{1903.09972}, the values of $ R_{cut} $ and $ L_{KT} $ were independent parameters of the fit, see Table~\ref{tab1}. For their ratio in the three considered scenarios, we obtain the values $ x = \{0.12,0.14,0.23 \} $, which coincide with the cosmological ratio in the order of magnitude.

Further, for the asymptotic orbital velocity, we obtain the expression

\begin{equation}
v_a^2=G(M_{lm}+M_{dm}(R_{cut}))/R_{cut}=G(1+x)M_{lm}/R_{cut}.
\end{equation}

The dependence of $ M_{lm} $ on $ R_{cut} $ has the form $ M_{lm} \sim R_{cut} ^ D $, where $ D = 3 $ for the classical homogeneous distribution. At the same time, in a number of papers \cite{Mandelbrot1997,9711073,0604496} it was noted that the observed distribution of luminous matter is better described by {\it fractal distribution} with the dimension $ D \sim2 $. From here we get Tully-Fisher relation:
\begin{eqnarray}
&&v_a^2\sim R_{cut}^{D-1}\sim M_{lm}^{(D-1)/D},\ M_{lm}\sim v_a^{2D/(D-1)},
\end{eqnarray}
whose slope is identical to that obtained by Kirillov and Turaev in their work \cite{0604496}:
\begin{eqnarray}
&&\beta=2D/(D-1),
\end{eqnarray}
as the dimension varies from $ D = 3 $ to $ D = 2 $, the slope changes from $ \beta = 3 $ to $ \beta = 4 $.

\section{T-asymmetric RDM solution: tachyonic oven}

Earlier, in works \cite{1701.01569,1707.02764,1811.03368,1812.11801,1903.09972} we considered stationary T-symmetric RDM solutions in which the incoming and outgoing flows were T-conjugated to each other. In this chapter we will relax this restriction and consider T-asymmetric flows, constrained only by the property of stationarity.

Consider the standard metric in spherical coordinates
\begin{equation}
ds^2=-A(r)dt^2+B(r)dr^2+r^2(d\theta^2+\sin^2\theta\; d\phi^2). \label{stdmetr}
\end{equation}
Select the energy-momentum tensor consisting of the sum over the flows
\begin{equation}
T^{\mu\nu}=\sum_i\rho_i(r)u_i^\mu(r)u_i^\nu(r),\label{Tmunu}
\end{equation}
where $ \rho_i (r) $ is the density and $ u_i (r) = (u_i ^ t (r), u_i ^ r (r), 0,0) $ are radially directed velocity fields for each flow. Note that there is no pressure in each flow, matter has a dust type, and flows pass through each other without interaction. Therefore, each flow obeys a geodesic equation, the general solution of which was written in \cite{1701.01569}, here we just have to choose own integration constants in each flow:
\begin{eqnarray}
&&4\pi\rho_i=c_{1i}/\left(r^2u_i^r\sqrt{AB}\right),\ u_i^t=c_{2i}/A, \ u_i^r=\sqrt{c_{2i}^2+c_{3i}A}/\sqrt{AB}.\label{eq_geode}
\end{eqnarray}
The first constant $ c_{1i}> 0 $ corresponds to positive density flows. For the second constant, in this paper, we consider both positive and negative values, which is equivalent to the incoming and outgoing flows $ \pm u_i ^ t (r) $ from \cite{1701.01569}. The third constant defines a norm $ c_{3i} = u_{i \mu} u_i ^ \mu $, for which three discrete options can be considered: $ c_{3i} = - 1 $ for massive, $ c_{3i} = $ 0 for null, $ c_{3i} = 1 $ for tachyon radial dark matter (M,N,T-RDM). Here, as well as in \cite{1811.03368}, we use $ \rho $ -equation with $ 4 \pi $ factor, corresponding to the choice of units $ G = c = 1 $.

The energy-momentum tensor is determined by the sums over the flows (or integrals for the continuous case, and in the specific numerical examples considered below, we consider two flows $ i = 1,2 $). The components of interest are
\begin{eqnarray}
&&T^{rr}=\sum_i\rho_i(r) (u_i^r)^2 = (4\pi)^{-1}r^{-2}A^{-1}B^{-1}\sum_i c_{1i}\sqrt{c_{2i}^2+c_{3i}A},\\
&&T^{tt}=\sum_i\rho_i(r) (u_i^t)^2 = (4\pi)^{-1}r^{-2}A^{-2}\sum_i c_{1i}c_{2i}^2/\sqrt{c_{2i}^2+c_{3i}A},\\
&&T^{tr}=\sum_i\rho_i(r) u_i^t u_i^r = (4\pi)^{-1}r^{-2}A^{-3/2}B^{-1/2}\sum_i c_{1i}c_{2i}=0,
\end{eqnarray}
here we have taken into account the condition for the energy balance of the flows $ T ^{tr} = 0 $, which becomes a condition for the constants $ \sum_i c_{1i} c_{2i} = 0 $, necessary for stationarity of the solution. The entire tensor takes the form $ T ^{\mu \nu} = \diag (T ^{tt}, T ^{rr}, 0,0) $, and the Einstein field equations (EFE) are written as
\begin{eqnarray}
&&rA'=-A(1-B) +8\pi AB^2r^2T^{rr},\\
&&rB'=B(1-B) +8\pi AB^2r^2T^{tt}.
\end{eqnarray}
After substitutions we get
\begin{eqnarray}
&&rA'=-A(1-B) +B\sum_i c_{4i}\sqrt{1+c_{5i}A},\\
&&rB'=B(1-B) +A^{-1}B^2\sum_i c_{4i}/\sqrt{1+c_{5i}A},\\
&&c_{4i}=2c_{1i}|c_{2i}|,\ c_{5i}=c_{3i}/c_{2i}^2,
\end{eqnarray}
or in logarithmic variables introduced in \cite{1701.01569}
\begin{eqnarray}
&&a=\log A,\ b=\log B,\ x=\log r,\\
&&da/dx=-1+e^b+e^{b-a}\sum_i c_{4i}\sqrt{1 + c_{5i}e^a},\\
&&db/dx=1-e^b+e^{b-a}\sum_i c_{4i}/\sqrt{1 + c_{5i}e^a}.
\end{eqnarray}

The procedure for the numerical integration of these equations was described in \cite{1701.01569}. Integration begins at large distances $ r_1 $, where the weak field conditions apply. At this point, by agreement, $ a_1 = 0 $ is set, corresponding to $ A_1 = 1 $, which is equivalent to measuring the global time $ t $ by the static observer's clock at this point. Due to the weak field conditions, the value of $ b_1 $ also turns out to be small, and the field equations in this mode can be linearized:
\begin{eqnarray}
&&da/dx=b+c_6,\ db/dx=-b+c_7,\label{eq_lin}\\
&&c_6=\sum_i c_{4i}\sqrt{1 + c_{5i}},\ c_7=\sum_i c_{4i}/\sqrt{1 + c_{5i}},
\end{eqnarray}
The general solution is
\begin{eqnarray}
&&a=Const+2\epsilon x -r_s e^{-x},\ b=c_7+r_s e^{-x},\ \epsilon=(c_6+c_7)/2.\label{eq_z1}
\end{eqnarray}

Identifying the value $ a / 2 = \varphi $ with the gravitational potential, we obtain the constant $ \epsilon = v_a ^ 2 $, which determines the asymptotic orbital velocity, and $ r_s = 2m $ -- the gravitational radius corresponding to the mass $ m $ .

\begin{figure}
\begin{center}
\includegraphics[width=\textwidth]{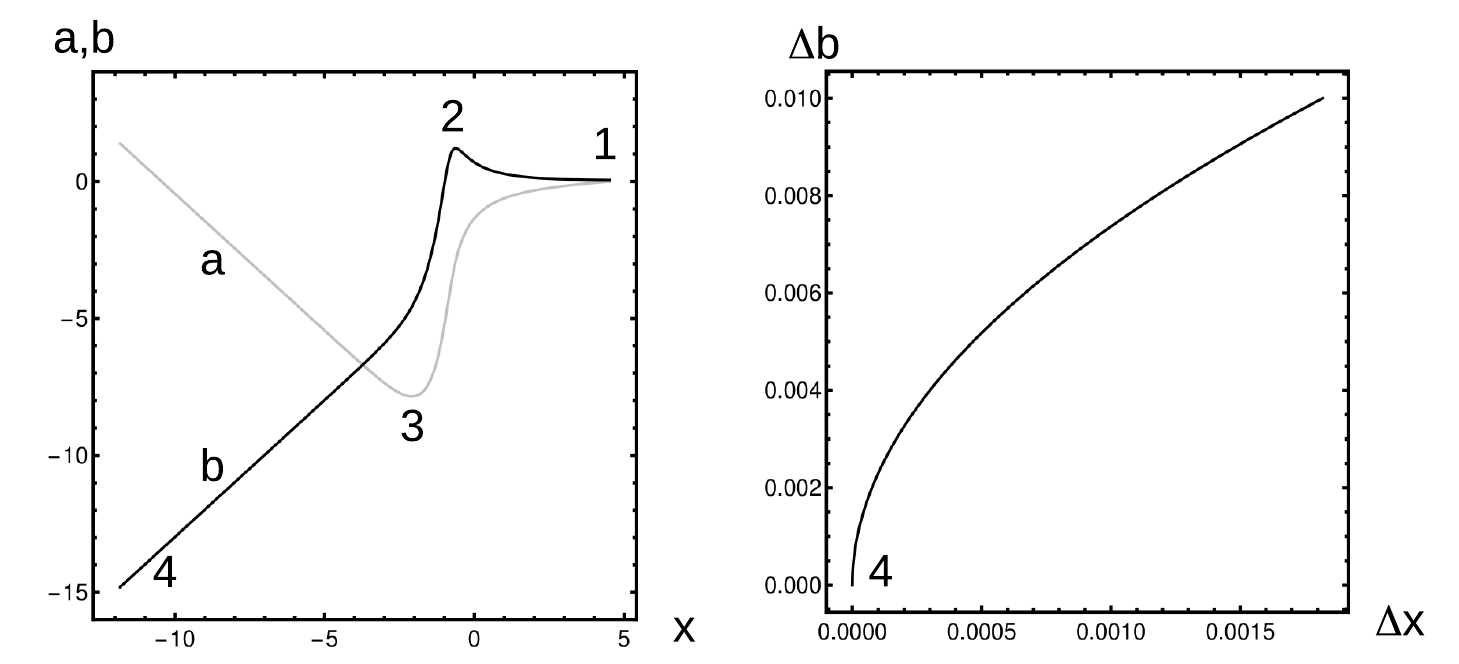}
\end{center}
\caption{Left: tachyonic oven solution in logarithmic coordinates. Right: a closeup shows the threshold corresponding to the turning point for the outgoing massive flow.}\label{f1}
\end{figure}

As a starting point for our further construction, we take the T-symmetric solution shown in \cite{1701.01569} in Fig.4, which is similar in shape to the one shown here in Fig.\ref{f1} left. The solution found has the following parameters
\begin{eqnarray}
&&\epsilon=0.04,\ c_2=\pm1,\ c_3=1,\\
&&c_1=\epsilon/2/(2c_2^2 + c_3)\sqrt{c_2^2 + c_3}=0.00942809.
\end{eqnarray}
There are two flows, incoming and outgoing, with the same mass and energy characteristics. Integration starts from point 1:
\begin{eqnarray}
&&r_s=1,\ r_1=100,\ a_1=0,\ b_1=c_7+r_s/r_1,
\end{eqnarray}
and goes further in the direction of decreasing $ r $. The constant $ c_7 $ is found from the selected model parameters using the above formulas.

Next, we determine the minimum of the $ a $-profile for this solution:
\begin{eqnarray}
&&a_3=-7.84795,\ b_3=-4.58068,\ x_3=-2.08905.\label{abx3}
\end{eqnarray}
After that, for verification, we start the integration from point 3 forward and backward in $ x $. This, of course, reproduces the same solution.

Now we modify the problem, making it T-asymmetric:
\begin{eqnarray}
&&c_{11}=0.00942809/5,\ c_{21}=-5,\ c_{31}=1,\\
&&c_{12}=0.00942809/2,\ c_{22}=2,\ c_{32}=-1,
\end{eqnarray}
for these two flows, $ T ^{tr} = 0 $ holds, the first flow is incoming tachyonic, the second is outgoing massive, the coefficient $ c_{2i} $ is also unbalanced between the flows. Like previously, the integration is started from point 3, setting the starting values to (\ref{abx3}). We obtain the solution shown in Fig.\ref{f1}.

Surprisingly, the solution practically does not change, only the cutoff appears in the left side, at point 4. The closeup in Fig.\ref{f1} right shows the threshold that we previously saw for massive T-symmetric solutions, \cite{1701.01569}. The massive flow is repelled from the central singularity of the negative mass and turns back. The solution cannot be continued to the left; a vacuole forms around the center. Now we have one massive flow, in this particular case -- the outgoing one. To interpret this solution, it is necessary to cut it at the minimum $ r $ or higher, e.g., placing the core there. On the surface of the core, the incoming tachyonic flow turns into the massive outgoing one (or vice versa).

The property that the solution is modified only a little is due to the fact that most of the solution has large negative values of the $ a $-profile. Due to this, the matter terms are disabled, $ c_{5i} e ^ a $ are small. Thus, the solution does not depend on $ c_{5i} $ and depends on the other coefficients only on the combination $ c_{4i} = 2c_{1i} | c_{2i} | $ summed over all the flows. When going in the direction of point 1, the solution goes to the limit of weak fields with small $ a, b $, which looks the same in the common scale, regardless of the matter terms. When going in the direction of point 4, the matter terms wake up, but there they are suppressed by the common factor $ e ^{b-a} \ll1 $. Therefore, the form of solution turns out to be close to the unperturbed one. However, if $ c_{5i} $ is negative and $ a $ is unlimitedly increasing, sooner or later the threshold $ 1 + c_{5i} e ^ a = 0 $ appears, and the solution cannot be continued further. This, however, is not related to the T-asymmetry of the solution, but is typical of any solutions containing radial flows of massive matter. Ultimately, it becomes clear that the consideration of T-asymmetric flows leads to the solutions close with T-symmetric ones also in the case of three or more flows. It is only required that they are energetically balanced.

\section{Limiting massive RDM solution: shell condensate}

\begin{figure}
\begin{center}
\includegraphics[width=\textwidth]{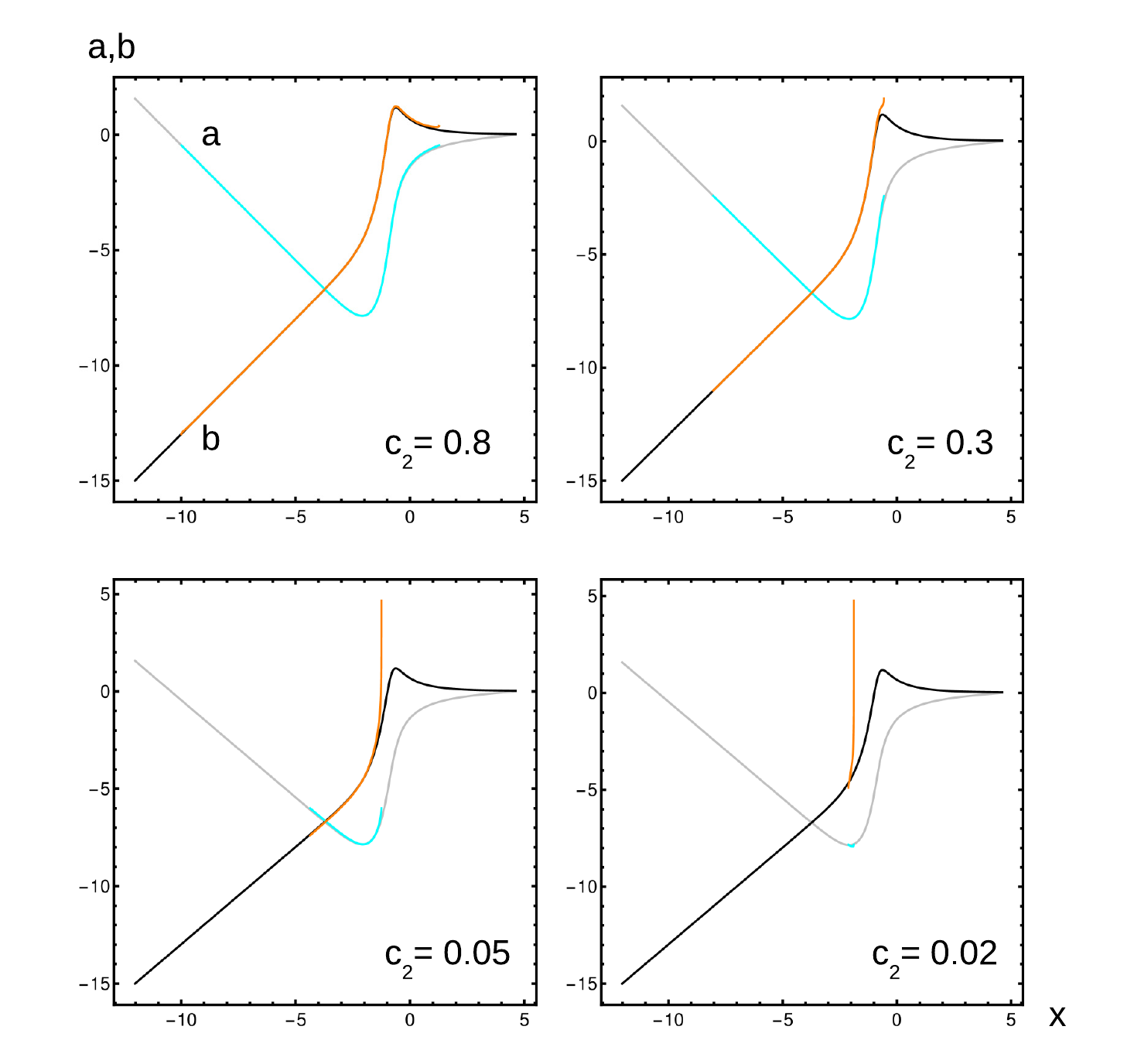}
\end{center}
\caption{Shell condensate solution (cyan-orange) on the background of the starting solution ($ c_2 = 1 $, black-gray).}\label{f2}
\end{figure}

Now we will again consider the 2-flow T-symmetric solution of massive type:
\begin{eqnarray}
&&c_{1i}=0.00942809/c_2,\ c_{2i}=\pm c_2,\ c_{3i}=-1,
\end{eqnarray}
we start the solution from point 3 according to the parameters (\ref{abx3}). Initially, $ c_2 = 1 $, then we lower the value of $ c_2 $. The result is shown in Fig.\ref{f2}. Since the expression $ U = c_2 ^ 2-e ^ a $ becomes negative for large $ a $ in the left and right sides of the solution, a potential well is formed in which the solution is locked. Turning points restrict the solution from above and from below. As $ c_2 $ decreases, the solution condenses to the bottom of the potential well. As a result, a thin massive shell is formed, in equilibrium in the repulsive field of the central negative mass. At the same time, the $ b $-profile becomes almost vertical. This behavior corresponds to the jump of Misner-Sharp mass
\begin{eqnarray}
&&M=r/2\ (1-B^{-1})=e^x/2\ (1-e^{-b}) 
\end{eqnarray}
from the negative mass inside to the positive mass outside. Outside the solution is continued as Schwarzschild solution with the positive mass.

\section{RDM solution with cosmological constant}

Inserting the cosmological constant in the Einstein equations
\begin{eqnarray}
&&G_{\mu\nu}= 8\pi T_{\mu\nu} - \Lambda g_{\mu\nu}, \label{EFELam}
\end{eqnarray}
RDM-equations are modified as follows:
\begin{eqnarray}
&&rA'=(A(1 - B^{-1}) + 2c_4 \sqrt{1 + c_5 A} -Ar^2 \Lambda) B,\label{EFELam1}\\  
&&rB'=(-A(1 - B^{-1}) + 2c_4 /\sqrt{1 + c_5 A} +Ar^2 \Lambda) B^2/A.\label{EFELam2}
\end{eqnarray}
These equations can be derived directly from (\ref{EFELam}). It is also interesting to trace their connection with the \cite{1707.02764} equations obtained for the combination of RDM and anisotropic gas, represented by the energy-momentum tensor $ T_ \mu ^ \nu = \diag (- \rho_{gas}, p_r, p_t, p_t) $. (Note also that here we use a different normalization of the coefficients, $ 2c_4 $ here corresponds to $ c_4 $ in \cite{1707.02764}.)

The cosmological term appears here as a special case
\begin{eqnarray}
&&\rho_{gas}=\Lambda/(8\pi),\ p_r=p_t=-\Lambda/(8\pi),
\end{eqnarray}
where all components are constant. The field equations from \cite{1707.02764} take the form (\ref{EFELam1}), (\ref{EFELam2}), while there is also a hydrostatic equation
\begin{eqnarray}
&&r (p_r+ \rho_{gas}) A'_r + 2 A\ (r (p_r)'_r + 2 p_r - 2 p_t)=0, \label{hydro}
\end{eqnarray}
which is satisfied identically after the substitutions. All other components of the Einstein equations are automatically satisfied on the surface (\ref{EFELam1}), (\ref{EFELam2}), (\ref{hydro}).

From the equations obtained, due to the smallness of the $ \Lambda $-term, it follows that its contribution begins to manifest itself only at very large distances from the center. For definiteness, consider the case of NRDM, $ c_5 = 0 $, $ 2c_4 = \epsilon $. Other types of matter produce similar result. The contribution of NRDM becomes comparable with the contribution of the $ \Lambda $-term at $ \epsilon \sim Ar ^ 2 \Lambda$. Note that most of the solution has very small $ A $, so the contribution of the $ \Lambda $-term in this zone is negligible. Near the naked singularity, $ A \sim Const / r $ holds, also with a very small constant, and the contribution of the $ \Lambda $-term in this zone is also negligible. Only at very large distances, in the weak field mode $ A \sim1 $, the required equality can be satisfied. Estimating the distances at which this occurs gives $ r_\Lambda \sim \sqrt{\epsilon / \Lambda} $. For the value of $ \Lambda = 3 (H_0 / c) ^ 2 \Omega_ \Lambda \sim10 ^{- 52} m ^{- 2} $ \cite{1807.06209} and for the MW value $ \epsilon = 4 \cdot10 ^{- 7} $, we obtain a theoretical estimate of $ r_\Lambda \sim2 $Mpc. In fact, much earlier, at distances of $ R_{cut} \sim50 $ kpc, the RDM model is cut off. At such distances, the contribution of the $ \Lambda $-term in relation to the DM-term in the equations is $ R_{cut} ^ 2 \Lambda / \epsilon \sim6 \cdot10 ^{- 4} $. Thus, throughout the range of the RDM model, when choosing the parameters corresponding to MW, the contribution of the cosmological constant to the equations is negligible.

\section{RDM solution on Penrose diagrams}

\begin{figure}
\begin{center}
\includegraphics[width=\textwidth]{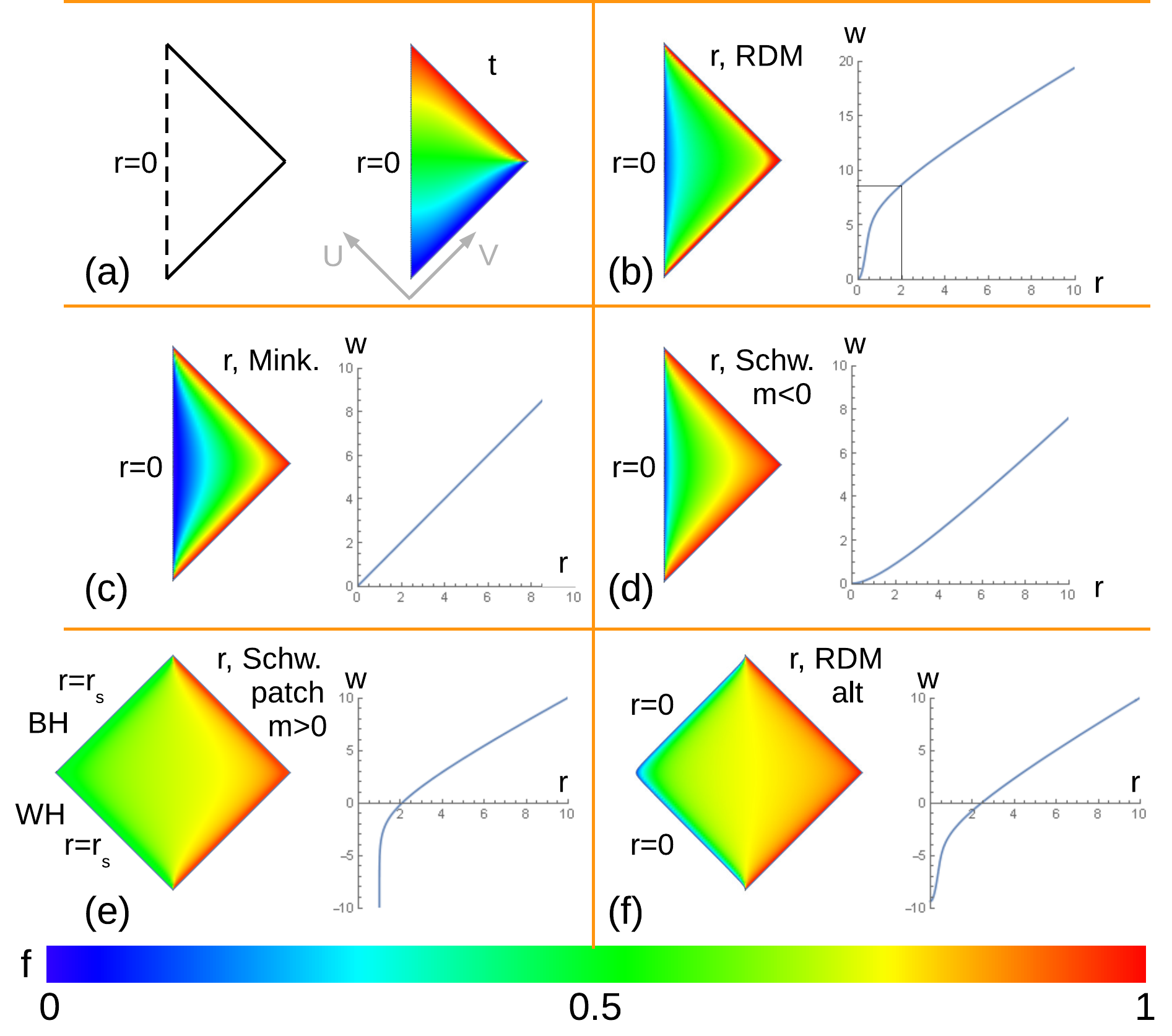}
\end{center}
\caption{Penrose diagrams for RDM model in comparison with flat space and Schwarzschild solutions of negative and positive mass.
}\label{f7}
\end{figure}

Consider the class of stationary spherically symmetric solutions with a metric of the form (\ref{stdmetr}). According to \cite{1701.01569}, $ A> 0 $, $ B> 0 $, the solution has no trapping horizons and has a naked singularity $ A \sim r ^{- 1} $, $ B \sim r $ at $ r \to0 $.

It is clear that the Penrose diagram is identical with the Schwarzschild solution for a mass $ m <0 $, which also has a naked singularity. The diagram has the form of a triangle in Fig.\ref{f7}a on the left. At the same time, at large distances, the solution behaves like the Schwarzschild one with $ m> 0 $, to which the radial flows of dark matter, also of positive mass, are coupled. These properties reflect themselves in the detailed structure of Penrose diagrams, for the analysis of which it is necessary to depict the fields of the temporal and radial coordinates $ t, r $ in the diagrams. To represent these fields, we use description \cite{Blau2018} and visualize the fields by color.

Radial light geodesics for the metric (\ref{stdmetr}) are given by the integral
\begin{eqnarray}
&&t=\pm w(r)+Const,\ w(r)=\int_0^r dr’ \sqrt{B(r’)/A(r’)},
\end{eqnarray}
the presence of an arbitrary constant reflects the property of stationarity, due to which the geodesics can be freely translated along the $t$ axis. The sign $ \pm $ reflects the T-symmetry property of solutions (outgoing / incoming geodesics). The asymptotics of the solution for $ r \to0 $ causes the integral in this formula to converge at the lower limit, that is, radial light geodesics reach a singularity in a finite time. The function $ w (r) $ is monotonically increasing with the initial value $ w (0) = 0 $.

Next, the double null coordinates $ u = t - w (r) $ and $ v = t + w (r) $ are introduced with the range $ - \infty <u \leq v <\infty $ and the compactified coordinates $ U = \arctan u $, $ V = \arctan v $, with the domain $ - \pi / 2 <U \leq V <\pi / 2 $. Thus, the region of variation of variables on the Penrose diagram is indeed a triangle, see Fig.\ref{f7}a on the left. In the case, if the integral $ w (r) $ diverges, as occurs, for example, for the Schwarzschild solution with $ m> 0 $ near the event horizon, radial light geodesics would have a range of $ - \infty <w (r ) <\infty $ and the variables domain would be a square $ - \pi / 2 <U, V <\pi / 2 $ or a rhombus in the rotated coordinates of Fig.\ref{f7}e.

Further, we note that the field $ t = (\tan U + \tan V) / 2 $ does not depend on the detailed structure of the metric and for all the models discussed below is represented by the color field in Fig.\ref{f7}a right. If necessary, the field continues to the entire rhombus with a reflection about the line $ r = 0 $. The field $ w (r) = (\tan V - \tan U) / 2 $ is also independent of the metric and is obtained from the field $t$ in the rhombus by turning it on $ 90 ^ \circ $, followed by a restriction to the triangle and an appropriate recoloring, see Fig.\ref{f7}c. Due to the monotony of $ w (r) $, the isolines of $ r = Const $ for all the models under consideration have the same form and differ only by relabeling of the values of $r$ assigned to them. In other words, the color fields encoding $r$ for different models are distinguished by a one-dimensional diffeomorphism of the color scale. The scale displays the value
\begin{eqnarray}
&&f=2/\pi\cdot\arctan(rc_{vis}),\ f=1/2+1/\pi\cdot\arctan(t),
\end{eqnarray}
$ f $ varies within $ [0,1] $, $ rc_{vis} = 1 $ and $ t = 0 $ correspond to $ f = 0.5 $. All examples considered below will be dimensioned to $ r_s = 2m = 0, \pm1 $, the constant $ c_ {vis} $ will be chosen conveniently for visualization, $ c_ {vis} = 5 $ for Fig.\ref{f7}b and $ c_ {vis} = 1 $ for the other figures.

Fig.\ref{f7}b on the right shows the outgoing radial light geodesic $ w (r) $ with the starting value $ w (0) = 0 $ for RDM solutions with $ \epsilon = 0.04 $, $ r_s = 1 $. The solution was found using numerical integration, see \cite{1701.01569} for details. The left part of this figure displays the corresponding $r$-field. Due to the choice of the constant $ c_ {vis} = 5 $, the rectangular area selected in Fig.\ref{f7}b right is actually visualized.

For Schwarzschild solutions
\begin{eqnarray}
&&A=B^{-1}=1-r_s/r,\ w=r + r_s \log(r-r_s)+const. \label{wschw}
\end{eqnarray}

For comparison, the following figures show:
\begin{itemize}
\item [(c)] -- the empty Minkowski space, $ r_s = 0 $, $ w (0) = 0 $, $ w = r $;
\item [(d)] -- Schwarzschild solution for $m<0$, $ r_s = -1 $, $ w (0) = 0 $, $ w = r- \log (r + 1) $;
\item [(e)] -- Schwarzschild patch for $m>0$, $ r_s = 1 $, $ w = r + \log (r-1) + const $;
\item [(f)] -- an alternative RDM solution with the same parameters, where $ w (r) $ is shifted by a constant.
\end{itemize}

The constants in solutions e,f are chosen in a way to achieve a match of $ w (r) $ functions with solution c at large distances. In the simplest way, this match can be obtained by requiring for solutions e,f, instead of $ w (0) = 0 $, the condition $ w (r_1) = r_1 $ for some large $ r_1 $. Although with a further increase in $r$, the solutions will again go apart due to the presence of logarithmic terms in them, these deviations will be small compared to the leading  $r$-term. Thus, the solutions will be close in a relative sense, which is enough for our color representation of the $r$-field. After the described matching procedure, the parts of c,e,f diagrams near the right edge look identically, representing the same behavior of the radial light rays as they move to infinity. 

When comparing the diagrams of Fig.\ref{f7}b and Fig.\ref{f7}d, the region near $ r = 0 $ has a similar structure, corresponding to the naked Schwarzschild singularity. Further, in Fig.\ref{f7}b there is a wide green region corresponding to the rapid variation of the $ w (r) $ function, the supershift. After this, a transition to the asymptotic mode occurs, $ w \sim r $ in the leading order. In Fig.\ref{f7}d, the supershift part is absent, the field immediately goes into the asymptotic mode.

When comparing the diagrams of Fig.\ref{f7}e and Fig.\ref{f7}f, the main difference is that $ r = r_s $ on the left border of the rhombus in Fig.\ref{f7}e, while in Fig.\ref{f7}f the left boundary corresponds to $ r = 0 $ and it is a smooth curve passing inside the rhombus. This curve is a timelike worldline of the naked Schwarzschild singularity, where at each point two radial light geodesics are terminated, one ingoing and one outgoing. While for the solution in Fig.\ref{f7}e, the left boundaries are the lightlike horizon lines. The horizons in this diagram are separated, the upper left edge corresponds to the event horizon of the black hole and only the incoming light rays are absorbed by it, while the lower left edge corresponds to the Cauchy horizon of the white hole and only the outgoing light rays emanate from it. Because of this, the solution in Fig.\ref{f7}e is geodesically incomplete, in particular, the light rays passing under $ \pm45 ^ \circ $ pierce the horizon $ r = r_s $ and can be continued by completing the Penrose diagram to the well-known Kruskal-Szekeres solution.

In the diagram in Fig.\ref{f7}f, one can also trace the transition to the limit $ \epsilon \to + 0 $ for the RDM solution. Keeping fixed the asymptotic structure of solution on the right border, at $ \epsilon \to + 0 $, the graph $ w (r) $ at $ r \leq r_s $ will deepen more and more into the negative region, turning into the Schwarzschild solution in Fig.\ref{f7}e. In this case, on the Penrose diagram, the line $ r = r_s $ and the left border $ r = 0 $ will be pressed more and more against the left edge of the diagram, in the limit the line $ r = 0 $ disappears and the diagram in Fig.\ref{f7}e with $ r = r_s $ on the left edge will be obtained.

Diagrams b and f are conformally equivalent. Due to the fact that their generators of geodesics are shifted by a constant, $ w (r) \to w (r) + c $, in the expression $ w (r) = (\tan V- \tan U) / 2 $ one can apply one-dimensional diffeomorphism $ V \to \arctan (\tan (V) + c) $ or $ U \to \arctan (\tan (U) -c) $ to achieve the same effect. This transformation preserves the form of the metric $\sim dUdV$.

We draw attention to the property noted in \cite {1701.01569,1811.03368,1812.11801} that the supershift occurs at values of $r$ less than the nominal value of $ r_s $. The effect becomes greater with the increasing $ \epsilon $, in particular, for $ \epsilon = 0.04 $, used here for plotting the diagrams, the supershift occurs at values of $r$ that are much smaller than $ r_s $. This fact is noticeable in the graphs of $ w (r) $ for RDM solutions. It also requires an increase in the constant $ c_ {vis} $ to provide convenient visualization of $r$-field. With $ \epsilon \to + 0 $, the supershift becomes stronger and occurs at values of $r$, closer to $ r_s $.

Another caveat needs to be made about the behavior of RDM solutions over long distances. At distances $ r \gg r_s $, the weak field theory begins to operate, in which the gravitational potential in geometrized units has the form $ \varphi = \epsilon \log (r / r_1) $, where $ 0 <\epsilon \ll1 $. The constant $ r_1 $ here has a gauge meaning, namely, the clock, by which the global time $t$ is measured, is located at the distance $ r_1 $ from the center. When analyzing a model with MW parameters, one can choose the distance $ r_1 = 100 $kpc on the outer border of the galaxy. At even greater distances, the potential reaches $ \varphi \sim1 $ and the field becomes large again. However, this occurs at very large distances $ r \sim r_1 \exp (1 / \epsilon) $, for MW $ \epsilon = 4 \cdot10 ^{- 7} $, $ r \sim 10 ^{10 ^ 6} $pc, which is much larger than the size of the observable universe. Practically, the solution will be cut off much earlier at the outer radius $ R_{cut} $ and joined to the cosmological model, as described before (however, this juncture is not shown here on Penrose diagrams). Also, for the parameter $ \epsilon = 0.04 $, which we use to construct the exemplary Penrose diagrams, the strong field regime is reached at $ r / r_1 \sim10 ^{11} $, which is realized only in a microscopically small area near the right border of the diagram. Outside this zone, for large $r$, the asymptotic regime is realized, close to the empty Minkowski space.

Summarizing the analysis of Penrose diagrams, for RDM solutions in Fig.\ref{f7}b,f we find similar elements on the diagrams for free space and Schwarzschild solutions for positive and negative masses. In this way, the RDM solution is combined of these elements.

\section{Origin of negative masses}

Now we make a general remark about the negative masses arising in the model under consideration. Although we introduce these masses phenomenologically, as the naked Schwarzschild singularity or incompressible core of negative density, their appearance may be due to the physical mechanism described in \cite{0602086,0604013,0607039,1401.6562,1409.1501}. The calculation within the loop quantum gravity \cite{0602086,0604013,0607039} shows that the field equations are modified when the Planck density is reached, so that the effective density is $ \rho_X = \rho \, (1- \rho / \rho_P) $. This density becomes zero at $ \rho = \rho_P $ and negative at $ \rho> \rho_P $, that is, when the Planck density is reached, the gravity turns itself off, and when it is exceeded, it turns into its opposite, the antigravity. In the works \cite{0602086,0604013,0607039} this effect was used to construct a cosmological scenario in which the compression of the universe ends at the finite non-zero radius and is followed by a {\it quantum bounce}, which replaces the singular Big Bang event. In the works \cite{1401.6562,1409.1501} this process was used in a miniature to describe {\it Planck stars}, black holes, which after reaching the Plank density by the collapsing matter, turn into white and eject this matter out. 

In the author's work \cite{1811.03368}, it was shown that the inclusion of negative mass core in the model resolves {\it Eardley instability} of white holes, which allows to construct stable models of white holes completely T-conjugated to black ones. Negative mass necessary for such stabilization can be obtained by the compression of the core to the densities exceeding the Planck one. The author's other work \cite{1812.11801} considered a stationary model with a frozen quantum bounce. In the center of the galaxy, inside the central RDM-star, there is a core with a density permanently exceeding the Planck one. As the estimate shows, a small relative excess of the density $ \Delta \rho / \rho_P \sim3 \epsilon = 1.2 \cdot10 ^{- 6} $ gives sufficient antigravity force created by the core to keep in hydrostatic equilibrium the dark matter halo supported by it. 

Thus, there is an interesting opportunity to create effectively the matter of negative mass for the needs of the model considered here, as well as for the other models ({\it warp drives} \cite{0009013}, {\it wormholes} \cite{Visser1996}, {\it time machines} \cite{Visser1996}), from ordinary matter, by compressing it to densities higher than the Planck one.

\section{Conclusion}

In this paper, a combined analysis of FRBs and RCs was performed within the framework of the RDM model. It was shown that the centers of spiral galaxies, close in structure to the Milky Way, can be sources of FRBs, also after the model parameters are constrained by observable RCs. The observed variation in the frequency of FRBs can be explained by the natural variability of the source parameters.

Within the framework of the RDM model, the Tully-Fisher relation was derived, with the slope parameter described by the Kirillov-Turaev formula $ \beta = 2D / (D-1) $, where $ D $ is the fractal dimension of the luminous matter distribution. When the dimension changes from $ D = 3 $ to $ D = 2 $, the slope changes from $ \beta = 3 $ to $ \beta = 4 $, with the experimentally observed $ \beta \sim3.6 $.

A number of particular solutions of the RDM model were considered. Tachyonic oven, a solution in which incoming tachyon flows are converted to massive outgoing ones. The example shows the existence of T-asymmetric solutions of the problem. The structure of these solutions is very close to the previously studied T-symmetric ones. This happens because the solution of field equations in the considered ultrarelativistic mode is determined by the total energy-momentum, which is almost independent on the individual matter characteristics of the flows. Also, for the stationarity of the solution, the energy balance of the flows is necessary, not their T-symmetry.

A shell condensate solution arises in the RDM model of the massive type with two turning points, in the limit when these two points approach each other. In this case, a stationary solution arises in the form of a thin shell of positive mass, which is in equilibrium in the repulsive field of the central negative mass. The total mass of this solution is positive, and it continues outwards as the standard Schwarzschild solution of positive mass.

The modification of the RDM solution was investigated when the cosmological constant is introduced into the equations. When choosing the parameters typical to the Milky Way galaxy, the solution becomes very close to the unmodified solution, up to the distances $ R_{cut} \sim50 $ kpc, on which the RDM solution is joined to the ambient cosmological model.

Penrose diagrams were constructed in the RDM model, whose structure turned out to be combined from the Penrose diagrams for Schwarzschild solutions of positive and negative mass.

The possibility of creating matter of negative mass due to the effects of quantum gravity, by compressing ordinary matter to densities higher than the Planck one, is discussed. Such matter can be used to create an antigravity core in the model of RDM-stars, as well as in the other models, including warp drives, wormholes and time machines.

\paragraph*{Acknowledgment.} Thanks to Kira Konich for proofreading the paper.

\end{document}